\documentclass[a4paper, amsfonts, amssymb, amsmath,  showkeys,superscriptaddress, nofootinbib, aps,  twocolumn, prl, floatfix]{revtex4-2}
\usepackage[english]{babel}
\usepackage{graphicx}
\usepackage{braket}
\usepackage{bm}
\usepackage{physics}
\usepackage{amsmath,amsfonts,amssymb}
\usepackage{xfrac}
\usepackage{dsfont}
\usepackage[colorlinks=true, linkcolor=blue, citecolor=blue, urlcolor=blue]{hyperref}
\usepackage[most]{tcolorbox}
\usepackage{comment}
\usepackage{comment}

\newcommand{\appropto}{\mathrel{\vcenter{
  \offinterlineskip\halign{\hfil$##$\cr
    \propto\cr\noalign{\kern2pt}\sim\cr\noalign{\kern-2pt}}}}}

\newcommand{\ro}[1]{{\color{red}#1}}

\newtcolorbox{blackblock}{
  enhanced, breakable, colback=white, colframe=black, coltitle=black, coltext=black,
  boxrule=0.5pt, arc=2mm, outer arc=2mm, arc is angular, top=2pt, bottom=2pt, left=4pt, right=4pt,
  before skip=10pt, after skip=10pt, fonttitle=\bfseries, title={}
}
\newtcolorbox{redblock}{
  enhanced, breakable, colback=red!5!white, colframe=red!75!black, coltitle=black, coltext=black,
  boxrule=0.5pt, arc=2mm, outer arc=2mm, arc is angular, top=2pt, bottom=2pt, left=4pt, right=4pt,
  before skip=10pt, after skip=10pt, fonttitle=\bfseries, title={}
}

\begin{document}
\title{A microscopic design rule for spin supersolids in triangular-lattice magnets}
%\title{Exchange anisotropy control as a design rule for spin supersolids}
%\title{Crystal-field control of supersolid--superfluid competition}
\author{Ryota Ono}
\email[Corresponding author: ]{ryota.ono.gm@gmail.com}
\affiliation{Advanced Science Research Center, Japan Atomic Energy Agency, 2-4 Shirakata, Tokai-mura, Ibaraki, 319-1195, Japan}

\author{Jun'ichi Ieda}
\affiliation{Advanced Science Research Center, Japan Atomic Energy Agency, 2-4 Shirakata, Tokai-mura, Ibaraki, 319-1195, Japan}
\author{Michiyasu Mori}
\affiliation{Advanced Science Research Center, Japan Atomic Energy Agency, 2-4 Shirakata, Tokai-mura, Ibaraki, 319-1195, Japan}
%\author{Igor Solovyev}
%\affiliation{National Institute for Materials Science, MANA, 1-1 Namiki, Tsukuba, Ibaraki 305-0044, Japan}
\author{Sadamichi Maekawa}
\affiliation{Advanced Science Research Center, Japan Atomic Energy Agency, 2-4 Shirakata, Tokai-mura, Ibaraki, 319-1195, Japan}
\affiliation{RIKEN Center for Emergent Matter Science (CEMS), Wako 351-0198, Japan}

\begin{abstract}
Spin supersolids emerge as a central topic in frustrated magnetism, motivating the search for realization in quantum materials. 
To this end, we study the origin of exchange anisotropy, $\Delta$, in triangular-lattice cobaltate families $X_2$$Y$Co(PO$_4$)$_2$ and $X_2$Co(SeO$_3$)$_2$ ($X$ = Na, K, Rb, Cs; $Y$ = Mg, Ca, Sr, Ba) by tailoring realistic spin models. 
We show that $\Delta$ is determined by the ratio of trigonal crystal field to spin–orbit coupling strength.
This framework explains contrasting anisotropies in these families, predicts systematic trends in $\Delta$ across $X/Y$-substitutions, and identifies candidate materials for spin supersolids. 
Our results establish trigonal field engineering as a microscopic route toward the design of spin supersolids.
\end{abstract}

\maketitle

Spin supersolids have attracted broad interest as an experimentally accessible realization of supersolidity in frustrated quantum magnets~\cite{10.1143/PTPS.46.411, Liu1973,PhysRevLett.95.127205,SSS_Kedar_PRL, PhysRevLett.95.127207,PhysRevLett.95.237204,Heidarian,PhysRevLett.112.127203,Huang_Maekawa}. 
They represent a rare quantum phase in which crystalline order and phase-coherent transverse magnetism coexist, while also offering promising low-temperature functionalities such as enhanced magnetocaloric response~\cite{Xiang2024} and potentially dissipationless spin transport~\cite{KonigPRL,PhysRevLett.112.227201,PhysRevLett.118.137201,Maekawa_SSS}. 
Because their stability depends sensitively on exchange anisotropy, understanding the microscopic origin of anisotropy is central to the materials design of spin supersolids~\cite{ulaga2025anisotropicheisenbergmodelclose}.

\begin{comment}
Spin supersolids have recently emerged as an experimentally accessible incarnation of supersolidity in frustrated quantum magnets~\cite{10.1143/PTPS.46.411, Liu1973,PhysRevLett.95.127205,SSS_Kedar_PRL, PhysRevLett.95.127207,PhysRevLett.95.237204,Heidarian,PhysRevLett.112.127203,Huang}.
They realize a coexistence of crystalline order and phase coherent transverse magnetism.
Here the “solid” component is a spatially modulated longitudinal magnetization that breaks lattice translational symmetry,
whereas the “superfluid” component is a transverse order parameter with a spontaneously chosen global phase, i.e., spontaneous breaking of U(1) spin rotation symmetry~\cite{10.1143/PTPS.46.411, Liu1973,PhysRevLett.95.127205,SSS_Kedar_PRL, PhysRevLett.95.127207,PhysRevLett.95.237204,Heidarian,PhysRevLett.112.127203,Huang}.
Such U(1) order typically relies on exchange anisotropy that reduces the SU(2) spin symmetry.
This makes spin supersolids unusually sensitive to the microscopic origin of exchange anisotropy in real materials~\cite{ulaga2025anisotropicheisenbergmodelclose}.
At the same time, they offer attractive functionalities, including an enhanced magnetocaloric response at low temperature~\cite{Xiang2024} and the prospect of dissipationless spin transport~\cite{KonigPRL,PhysRevLett.112.227201,PhysRevLett.118.137201,Maekawa_SSS}.
%Related supersolid-like magnetocaloric behavior has also been discussed in the chromium spinel MnCr$_2$S$_4$.
\ro{shorten spin supersolid?}
\end{comment}

Recent experiments have reported signatures of a spin supersolid in 
%two Co$^{2+}$-based triangular-lattice antiferromagnets (TLAFs), 
Na$_2$BaCo(PO$_4$)$_2$~\cite{Xiang2024,gsk8-1k9q} and K$_2$Co(SeO$_3$)$_2$~\cite{Zhu_PRL,KCSO_npj,Chen2026}.
In addition, for the newly synthesized Rb$_2$Co(SeO$_3$)$_2$, recent studies contradict each other on the emergence of the spin supersolid~\cite{RCSO1,RCSO2}.
They are members of the $X_2$$Y$Co(PO$_4$)$_2$ ({\it XY}CP) and $X_2$Co(SeO$_3$)$_2$ ({\it X}CSO) families ($X$=Na, K, Rb, Cs, $Y$ = Mg, Ca, Sr, Ba)~\cite{pnas.1906483116,PhysRevMaterials.4.084406,Li2020}, in which Co$^{2+}$ in distorted CoO$_6$ octahedra hosts a spin-orbit-entangled Kramers doublet on a quasi-two-dimensional triangular-lattice.
%Despite their similar local coordination, the experimental situations are not identical as follows \ro{$\Delta$?}.
%In NaBaCP, spin supersolid phases are supported rather robustly at both zero and finite magnetic fields~\cite{PNAS_NBCP,Xiang2024,gsk8-1k9q} \ro{neutron?}. 
%In KCSO, \bl{by contrast}, a zero-field supersolid appears plausible and a finite-field spin supersolid has been proposed~\cite{Zhu_PRL,KCSO_npj,Chen2026}  
%\ro{plausible, }, although the interpretation remains less settled than in NaBaCP.
%In NaBaCP, experiments consistently support supersolid phases in zero and finite magnetic fields, including phases below and above the one-third magnetization plateau~\cite{PNAS_NBCP,Xiang2024,gsk8-1k9q}.
%In KCSO, zero-field neutron spectroscopy and thermodynamic measurements have been interpreted in terms of a supersolid ground state~\cite{Zhu_PRL}, and a high-field supersolid adjacent to the one-third plateau has also been reported~\cite{KCSO_npj,Chen2026}.
%At the same time, because KCSO lies much closer to the Ising limit than NaBaCP, the microscopic interpretation of its supersolid signatures remains under more active discussion.
%While the experiments support three-sublattice solid order and the presence of a gapless excitation~\cite{Chen2026}, whether this excitation is the Goldstone mode expected for a supersolid remains an open question~\cite{PhysRevB.110.L180404,PhysRevB.110.214408}.
A widely used minimal model to
describe these materials is the pseudospin-$1/2$ XXZ model on a triangular-lattice 
\begin{align}
\mathcal{H}_{\rm XXZ}=\sum_{\langle ij\rangle} J \Big[
\left(e_i^x e_j^x+e_i^y e_j^y\right)+\Delta e_i^z e_j^z
\Big],
\label{eq:XXZ}
\end{align}
where $J>0$ is the antiferromagnetic exchange parameter, $\mathbf e_i=(e_i^x,e_i^y,e_i^z)$ denotes the pseudospin-$1/2$ operator at site $i$, and $\Delta$ is the exchange anisotropy~\cite{doi:10.1143/JPSJ.61.3732,PhysRevLett.114.027201,PhysRevB.91.081104}.
Theoretically, this model has been studied extensively as a function of $\Delta$~\cite{doi:10.1143/JPSJ.61.3732,PhysRevB.91.081104,PhysRevLett.114.027201,ulaga2025anisotropicheisenbergmodelclose,Huang_Maekawa}.
These studies establish that a supersolid is stabilized for $\Delta>1$, while a superfluid is stabilized for $\Delta<1$.
%Thus, controlling $\Delta$ provides a direct route to switch between supersolid and superfluid.
%Therefore, it is important to reveal what essentially controls $\Delta$ and how to control it chemically in real materials.

For materials design, however, one must identify which microscopic ingredient in the local electronic structure controls $\Delta$.
The issue is already evident experimentally: NaBaCP and KCSO have similar local CoO$_6$ coordination, yet their estimated anisotropies differ by almost an order of magnitude, $\Delta\simeq1.7$ and $\Delta\simeq14.3$, respectively~\cite{PNAS_NBCP,Gao2022,KCSO_npj}.
This striking contrast raises a question: what local ingredient controls $\Delta$, and can it be used to chemically drive a transition between supersolid and superfluid?

%On the other hand, a recent exact diagonalization study found that the spin supersolidity vanishes in the near-Ising regime at around $\Delta \sim 5$~\cite{ulaga2025anisotropicheisenbergmodelclose}.
\ro{
%On the other hand, the Ising limit ($\Delta \rightarrow \infty$) is known to exhibit the classical spin liquid.
%Therefore it is important to design $\Delta$ at chemical level.
}
%These developments make it essential to determine microscopically which anisotropy regime is realized in each material.
%Yet this task is hindered by the still limited understanding of how the local chemical environment controls $\Delta$.
%\ro{Therefore, emergence of the spin supersolid is extremely sensitive to $\Delta$, and the chemical control of it is important toward realization of spin supersolid.}
%Because these materials are Mott insulators, resolving this question requires a microscopic strong-coupling derivation based on superexchange theory, rather than conventional band theory approaches~\cite{Gao2022}.

In this Letter, starting from first-principles electronic structures and strong-coupling theory, we calculate anisotropic exchange parameters for the {\it XY}CP and {\it X}CSO families.
We further establish a symmetry-based mapping from the local crystal field to $\Delta$, providing a unified interpretation of our microscopic results.
This mapping quantitatively captures the overall trend of $\Delta$ across Co$^{2+}$-based triangular-lattice antiferromagnets (TLAFs) and provides a concrete route toward materials design of spin supersolids.

\begin{figure}[htbp]
\centering
\includegraphics[keepaspectratio, scale=0.11]{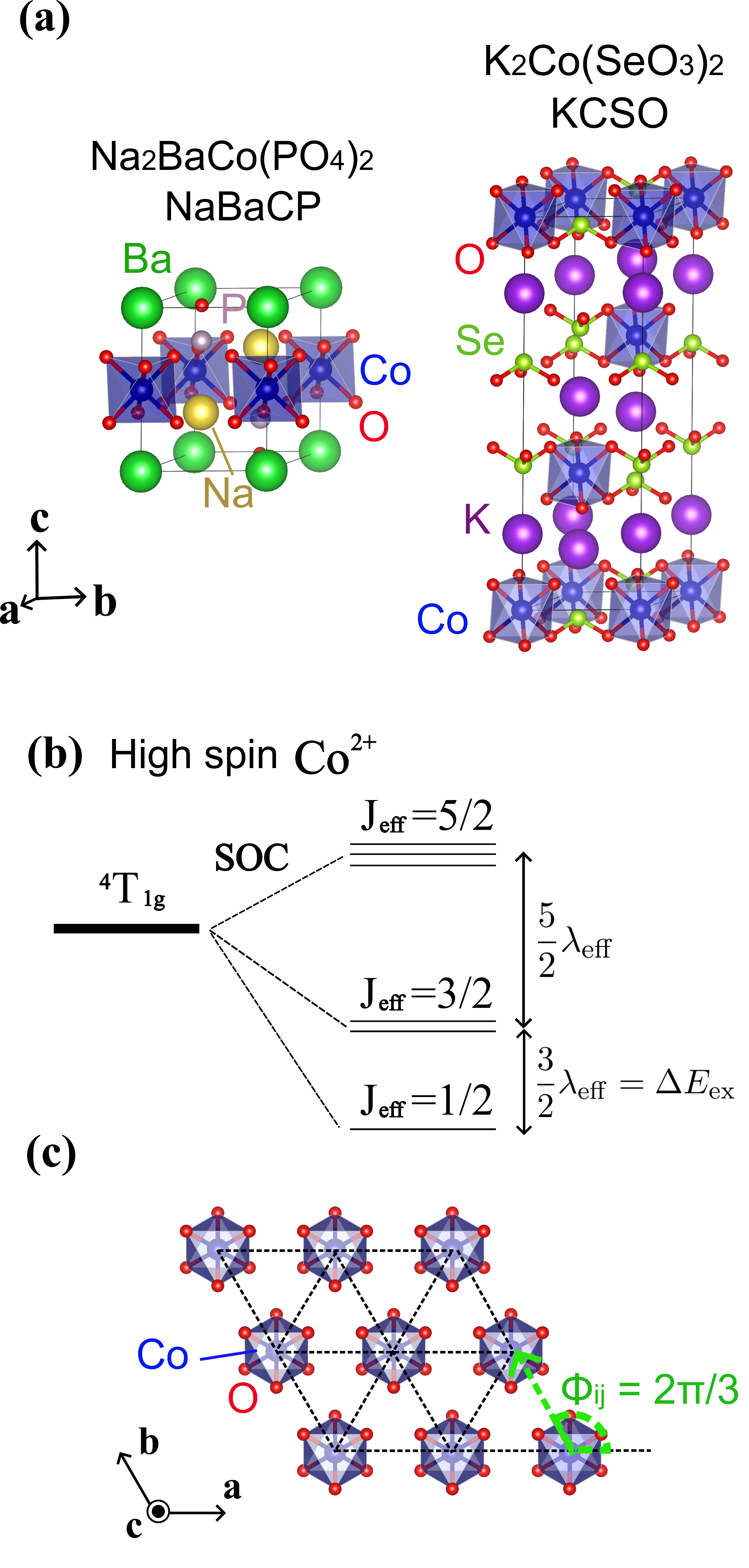}
\caption{
(a) Crystal structure of NaBaCP and KCSO.
(b) Energy level scheme of the lowest multiplet of Co$^{2+}$ in an octahedral crystal field with spin-orbit coupling (SOC). The energy gap between the $J_{\rm eff}=1/2$ ground state doublet and the first excited spin-orbit multiplet is denoted by $\Delta E_{\rm ex}$.
(c) triangular-lattice plane formed by CoO$_6$ clusters in NaBaCP and KCSO. The green dashed arrow indicates the bond along the crystallographic $\mathbf{b}$-axis ($\phi_{ij}=2\pi/3$).
}\label{fig:Co_Crystal}
\end{figure}

NaBaCP and KCSO crystallize in the space groups $P\bar{3}m1$~\cite{NBCP_crys,NBNP_NBCP_crys} and $R\bar{3}m$~\cite{KCSO_crys}, respectively [Fig.~\ref{fig:Co_Crystal}(a)].
In both cases, Co$^{2+}$ having pseudospin-1/2 Kramers doublet [$J_{\rm eff} = 1/2$ as shown in Fig.~\ref{fig:Co_Crystal}(b)] forms a quasi-two-dimensional triangular-lattice [Fig.~\ref{fig:Co_Crystal}(c)], implying inversion centers at the midpoints of all bonds.
%A notable difference is the interlayer stacking: NaBaCP exhibits $AA$ stacking of the triangular layers, whereas KCSO follows an $ABC$ stacking.
Moreover, the Co$^{2+}$ site has local $D_{3d}$ symmetry (threefold rotation with inversion).
Hypothetical substituted members of the {\it XY}CP and {\it X}CSO families are generated by structural optimization starting from the parent NaBaCP and KCSO structures.
These substitution provide a controlled way to tune the crystal field.
%\ro{The key leading crystal field after the usual octahedral one is the trigonal one.}

To study a relation between the crystal field and $\Delta$,
we construct a multi-orbital Hubbard model using electronic structures from first-principles calculations within generalized gradient approximation~\cite{GGA,QE1,SM},
%\begin{align}\label{eq:Hubbard}
%\mathcal{H}^{\rm el}=\mathcal{H}_{\rm one}+\mathcal{H}_U,
%\end{align}
in a spinor Wannier basis of five Co-$3d$ orbitals (ten spin-orbitals) obtained by the maximally localized Wannier-function method~\cite{Marzari,Marzari_mod}.
%The one-electron term is
%\begin{align}
%\mathcal{H}_{\rm one}
%=\sum_{ij}\sum_{ab}\sum_{\sigma\sigma'}
%h_{ij}^{ab;\sigma\sigma'}\, \hat c_{ia\sigma}^\dagger \hat c_{jb\sigma'},
%\end{align}
%where $\hat{c}_{ia\sigma}^{\dagger}$ ($\hat{c}_{ia\sigma}$) creates (annihilates) an electron in orbital $a$ with spin $\sigma$ on site $i$.
%The on-site block ($i=j$) contains the crystal field and atomic spin-orbit coupling, while the intersite block ($i\neq j$) defines the hopping integrals $t_{ij}^{ab;\sigma\sigma'}\equiv h_{ij}^{ab;\sigma\sigma'}$.
The screened on-site Coulomb interaction is evaluated by constrained random-phase approximation~\cite{Ferdi} and is represented in the spherical Kanamori form (see Sec. S1 of Supplemental Material (SM)~\cite{SM} for details of the electronic model).
%\begin{align}
%\mathcal{H}_{{\rm U},i}
%&= U \sum_a \hat{n}_{ia\uparrow}\hat{n}_{ia\downarrow}
%+ U' \sum_{a>b} \hat{n}_{ia}\hat{n}_{ib} \nonumber \\
%&- J_H \sum_{a>b}\sum_{\sigma\sigma'} \hat{c}_{ia\sigma}^\dagger \hat{c}_{ib\sigma'}^\dagger \hat{c}_{ia\sigma'} \hat{c}_{ib\sigma} \nonumber \\
%&+ J_H \sum_{a>b}\left(\hat{c}_{ia\uparrow}^\dagger \hat{c}_{ia\downarrow}^\dagger \hat{c}_{ib\downarrow}\hat{c}_{ib\uparrow} + \text{h.c.}\right).
%\end{align}
The resulting parameters and the detailed parametrization for {\it X}BaCP and {\it X}CSO are summarized in the Sec.~S1 of SM~\cite{SM}. 
The hopping integrals are found much smaller than the on-site Coulomb interaction, this justifies a strong-coupling superexchange treatment.

To extract realistic anisotropic exchange interactions, we perform a strong-coupling superexchange expansion of the Hubbard model~\cite{Anderson_SE,Solovyev_2008_SE,Solovyev_2009_SE,SE_Sergey,PRB_RIS,SM}.
First, we obtain the local $d^7$ ground-state Kramers doublet $|\Psi^\pm\rangle$ on each Co$^{2+}$ site.
Then, the Kramers doublet wave-function projected on to the pseudospin direction, $\pm a$ $(a=x,y,z)$, $\ket{\varphi^{\pm a}}$ is obtained as a linear combination $\ket{\varphi^{\pm a}}=c_{+}^{\pm a}|\Psi^+\rangle + c_{-}^{\pm a}|\Psi^-\rangle$, where $c_{\pm}^{\pm a}$ are linear combination coefficients of a Krmaers doublet.
Integrating out virtual charge fluctuations using this basis yield the second-order energy correction.
For a two-site pseudospin wave-function $\ket{\varphi^{\pm a}_i,\varphi^{\pm b}_j}$, we have:
\begin{align}
E_{ij}^{(2)}(\pm a,\pm b)
= \mel{\varphi^{\pm a}_i,\varphi^{\pm b}_j}{\hat{\mathcal{T}}_{ij}+\hat{\mathcal{T}}_{ji}}{\varphi^{\pm a}_i,\varphi^{\pm b}_j},
\end{align}
where $\hat{\mathcal{T}}_{ij}$ is the second-order perturbation operator associated with virtual electron hopping processes from site $j$ to $i$ (see Sec.~S2 of the SM for details of $\hat{\mathcal T}_{ij}$~\cite{SM}).
From the four collinear configurations, the exchange tensor is obtained as 
$
    J_{ij}^{ab} =[ E_{ij}^{(2)} (+ a,+ b) + E_{ij}^{(2)} (- a,- b) - E_{ij}^{(2)} (+ a,- b) - E_{ij}^{(2)} (- a,+ b)]/4
$
which yields the general bilinear pseudospin model
\begin{align}
\mathcal{H}=\sum_{\langle ij\rangle}\sum_{a,b=x,y,z} J_{ij}^{ab}\, e_i^{a} e_j^{b}.
\label{eq:Jani}
\end{align}
For the TLAF geometry considered here, all bonds are centrosymmetric and the Co site has local $D_{3d}$ symmetry, which forbids antisymmetric exchange and constrains $J_{ij}^{ab}$ to four independent parameters.
For a bond with in-plane angle $\phi_{ij}$ measured from the crystallographic $\mathbf{a}$-axis, the symmetry constrains the bond angle dependence as~\cite{TLAF_Jab,TLAF_Jab_PRB,TLAF_Jab_PRL2}
\begin{align}
&[J_{ij}^{ab}] \nonumber \\
&=\begin{pmatrix}
J+2J^{\mathrm{PD}}\cos(\phi_{ij}) & -2J^{\mathrm{PD}}\sin(\phi_{ij}) & -J^{\Gamma}\sin(\phi_{ij})\\
-2J^{\mathrm{PD}}\sin(\phi_{ij}) & J-2J^{\mathrm{PD}}\cos(\phi_{ij}) & \ \ J^{\Gamma}\cos(\phi_{ij})\\
-\,J^{\Gamma}\sin(\phi_{ij}) & \ \ J^{\Gamma}\cos(\phi_{ij}) & \Delta J
\end{pmatrix}.
\label{eq:Jtensor_D3d}
\end{align}
Here, for example, the bond along crystal $\mathbf{b}$-axis corresponds to $\phi_{ij}=2\pi/3$ [see a green dashed arrow in Fig.~\ref{fig:Co_Crystal}(c)].
The bond angle dependence arises solely from $J^{\mathrm{PD}}$ and $J^{\Gamma}$.
The XXZ limit Eq.~(\ref{eq:XXZ}) is recovered for $J^{\mathrm{PD}}=J^{\Gamma}=0$.
Our calculations show that these parameters are very small in the target materials.
We therefore neglect these non-XXZ components in the following discussion.
Such terms can nevertheless be important for bond dependent exchange parameters, such as those in the Kitaev model~\cite{PhysRevLett.102.017205, PhysRevLett.112.077204, Winter_2017}.
Although finite $J^{\mathrm{PD}}$ and $J^{\Gamma}$ may prevent true superfluidity at zero-temperature, it has been shown that finite temperature can stabilize it~\cite{JKKN,park2026spinorbitinducedinstabilityfinitetemperaturestabilization}.

\begin{figure*}[htbp]
\centering
\includegraphics[keepaspectratio, scale=0.11]{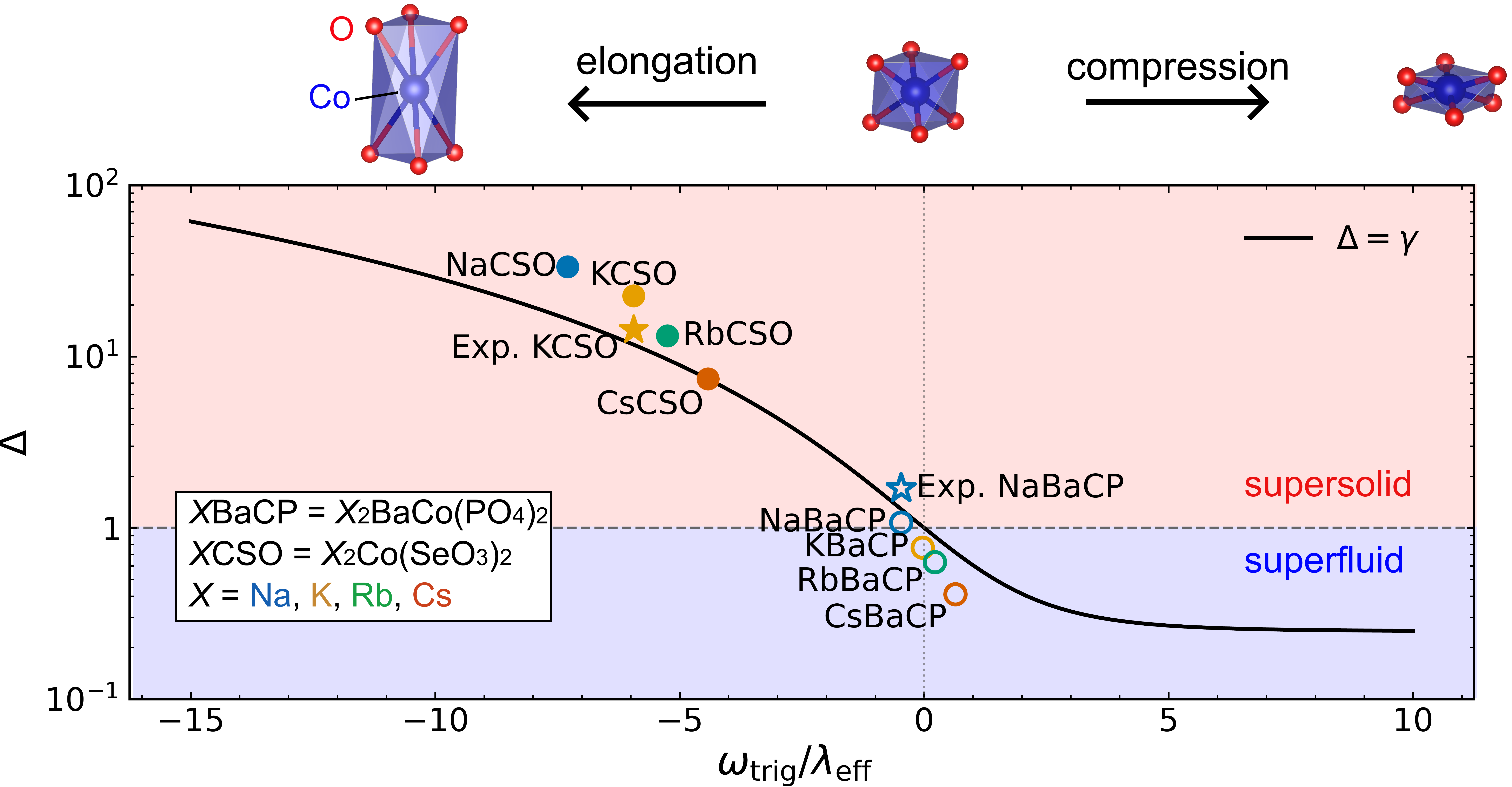}
\caption{
Plot of $\Delta$ as a function of $\omega_{\rm trig}/\lambda_{\rm eff}$ for the target materials. Solid and open circles denote calculated values for the {\it X}BaCP and {\it X}CSO series, respectively.  While solid and open stars indicate experimental estimates for KCSO and NaBaCP, respectively. The solid black curve shows the theoretical relation $\Delta=\gamma$ as a function of $\omega_{\rm trig}/\lambda_{\rm eff}$ obtained by solving the minimal single-ion model in Eq.~(\ref{eq:ion}).
Above the plot, schematic illustrations show the relation between the trigonal crystal field strength and the local CoO$_6$ distortion in a simplified picture that retains only the contribution of the surrounding O$^{2-}$ ions.
%The results under uniaxial compression along crystallographic $c$ axis in CsCSO are given in the inset. 
} \label{fig:crystal_alpha}
\end{figure*}

The nearest-neighbor (NN) exchange parameters of the $X$-site dependence for {\it X}BaCP and {\it X}CSO families are summarized in Fig.~\ref{fig:crystal_alpha}.
Star points in the Fig.~\ref{fig:crystal_alpha} indicate the experimental estimates of $\Delta$ (where available).
In both families the interlayer exchange parameters are negligible, and further neighbor ones are also small (see Sec.~S3 of SM for the full set of exchange parameters~\cite{SM}).
We therefore treat these compounds as effectively two-dimensional TLAFs.
For NaBaCP and KCSO, our exchange parameters are in qualitative agreement with estimates from linear spin-wave fits~\cite{PNAS_NBCP,KCSO_npj}.
Notably, $\Delta$ differs strongly between {\it X}BaCP and {\it X}CSO families, and in both series $\Delta$ decreases systematically upon substituting $X$ from Na to Cs.
%By contrast, KBaCP, RbBaCP, and CsBaCP lie in $\Delta < 1$ and are therefore expected to host superfluid rather than supersolid.
The calculated $\Delta$ already identifies several materials beyond NaBaCP and KCSO as supersolid candidates.
At the same time, the {\it X}BaCP series shows that $X$-site substitution can tune the XXZ model across the $\Delta=1$ boundary, as KBaCP, RbBaCP, and CsBaCP move to the superfluid side.

\begin{figure}[htbp]
\centering
\includegraphics[keepaspectratio, scale=0.65]{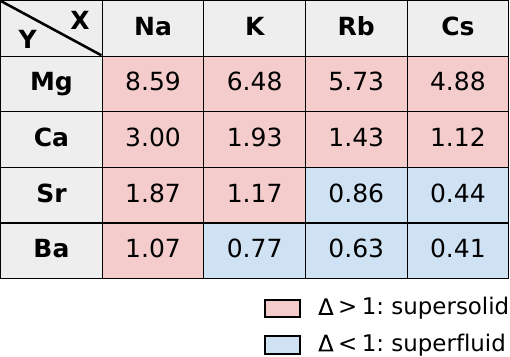}
\caption{ $\Delta$ in the {\it XY}CP family by combined $X$- and $Y$-site substitutions.
The red and blue colors denote $\Delta>1$ and $\Delta<1$, respectively, corresponding to the supersolid and superfluid of the XXZ phase diagram.
}
\label{fig:ABCP}
\end{figure}

%Finally, having identified the oxygen mediated chemical pressure mechanism, we extend it to combined $A$- and $B$-site substitutions in the {\it AB}CP family.
Furthermore, Fig.~\ref{fig:ABCP} presents the map of $\Delta$ in the $X$--$Y$ composition space in the {\it XY}CP family.
%The same structural mechanism operates throughout this map. 
%Replacing either the $A$- or $B$-site ion with a smaller cation enhances the $c$-axis dominated relaxation of the CoO$_6$ environment, shifts $\omega_{\rm trig}$ toward more negative values, and thereby increases $\Delta$.
Except for the already known NaBaCP member, the calculated compounds in the $\Delta>1$ (red) region represent previously unexplored {\it XY}CP supersolid candidates.
In particular, the Mg- and Ca-based {\it XY}CP families remain robustly on the supersolid for all $X$-site considered here.
The Sr- and Ba-based families, by contrast, approach or cross the $\Delta=1$ boundary depending on the $X$-site, demonstrating that the same mechanism can tune the system between the supersolid and superfluid.
Similarly, Na-based Na{\it Y}CP family also remains on the supersolid for all $Y$-site considered here.
%Thus, combined $A$- and $B$-site substitutions provide a microscopic route for designing supersolids by controlling the trigonal field.

\begin{comment}
\begin{table}[htpb]
\caption{Spin model parameters (meV) for $A_2$BaCo(PO$_4$)$_2$ ({\it A}BaCP) and
$A_2$Co(SeO$_3$)$_2$ ({\it A}CSO), $A=$ Na, K, Rb, Cs.
Values in parentheses are from experimental linear spin-wave fits (where available).
%Representative ranges of $J^{\rm PD}$ and $J^\Gamma$ are shown as extra rows. 
Full values including $J^{\rm PD}$ and $J^\Gamma$ are listed in Table~S1 of SM~\cite{SM}.}
\label{tab:param}
\begin{ruledtabular}
\setlength{\tabcolsep}{4.0pt}
\renewcommand{\arraystretch}{0.95}
\begin{tabular}{lcccc}
$A$ & $J$ & $\Delta$ & $\gamma$ & $\omega_{\rm trig}/\lambda_{\rm eff}$ \\
\hline
\multicolumn{5}{l}{{\it A}BaCP: $A_2$BaCo(PO$_4$)$_2$}\\
Na & 0.132\,(0.076)  & 1.1 (1.7) & 1.20 & -0.47 \\
K  & 0.127           & 0.8       & 0.85 & -0.03 \\
Rb & 0.123           & 0.6       & 0.71 &  0.22 \\
Cs & 0.118           & 0.4       & 0.49 &  0.64 \\
%\multicolumn{6}{l}{\hspace{1.2em} \footnotesize $J^{\rm PD}=-0.010\ldots-0.009$ meV,\ $J^\Gamma=-0.006\ldots-0.003$ meV}\\
\hline
\multicolumn{5}{l}{{\it A}CSO: $A_2$Co(SeO$_3$)$_2$}\\
Na & 0.056           & 33.3        & 34.90 & -7.29 \\
K  & 0.084\,(0.217)  & 22.6 (14.3) & 23.62 & -5.94 \\
Rb & 0.126           & 13.2        & 13.70 & -5.25 \\
Cs & 0.203           & 7.4         & 7.63  & -4.42 \\
%\multicolumn{6}{l}{\hspace{1.2em} \footnotesize $J^{\rm PD}=-0.001\ldots 0$ meV,\ $J^\Gamma=0.003\ldots0.010$ meV}\\
\end{tabular}
\end{ruledtabular}
\end{table}
\end{comment}

To understand the microscopic origin of $\Delta$, we first isolate the single-ion contributions to it in {\it XY}CP and {\it X}CSO.
In an octahedral environment, the lowest-energy term of high-spin Co$^{2+}$ is $^{4}T_{1g}$~\cite{Tanabe-Sugano}, which can be represented by an effective orbital moment $L_{\rm eff}=1$ coupled to $S=3/2$.
Atomic SOC splits this manifold into Kramers degenerate multiplets, with a $J_{\rm eff}=1/2$ Kramers doublet at the ground state [Fig.~\ref{fig:Co_Crystal}(b)].
The leading deviation from the octahedral symmetry allowed by the TLAF local geometry is an axial trigonal crystal field.
We thus consider the minimal single-ion Hamiltonian
\begin{align}\label{eq:ion}
\mathcal{H}_{\rm ion}= \omega_{\rm trig}[(L_{\rm eff}^z)^2-2/3]+\lambda_{\rm eff}\,\mathbf{L}_{\rm eff}\cdot\mathbf{S},
\end{align}
and note that $J_z$ is conserved (since $[J_z,L_z^2]=0$).
Here the first term represents the trigonal crystal field ($\omega_{\rm trig}$) and the second term represents atomic SOC ($\lambda_{\rm eff} > 0$).
Diagonalizing this Hamiltonian, we find the ground state Kramers doublet as
\begin{align}
|\psi_{+}\rangle &= c_1|+1,-\tfrac{1}{2}\rangle+c_2|0,+\tfrac{1}{2}\rangle+c_3|{-1},+\tfrac{3}{2}\rangle, \\
|\psi_{-}\rangle &= \hat{T}|\psi_{+}\rangle,
\end{align}
in the $\ket{L_{\rm eff}^z,S^z}$ basis, where $\hat{T}$ is the time-reversal operator. 
In this basis the Hamiltonian can be chosen real, so we take $c_1,c_2,c_3$ to be real without loss of generality.
We then introduce the spin projection factor $s^a$ ($a=x,y,z$) for a generic Kramers doublet system (see Sec.~S4g of the SM for details~\cite{SM}). 
Within this doublet, the spin projection factors are~\cite{Oguchi,Menshikov2014}
\begin{align}\label{eq:s}
s^z&= (-c_1^{2}+c_2^{2}+3c_3^{2})/2, \\
s^{x}&=s^{y} = \left|\sqrt{3}c_1c_3+c_2^{2}\right|  \equiv s^{\perp}.
\end{align}
We then define the single-ion anisotropy factor
\begin{align}
\gamma \equiv \left(\frac{s^{z}}{s^{\perp}}\right)^2 =  \left( \frac{-c_1^{2}+c_2^{2}+3c_3^{2}}{2 \left|\sqrt{3}c_1c_3+c_2^{2}\right|}\right)^2.
\label{eq:gamma_def}
\end{align}
$\gamma$ becomes a universal function of a dimensionless ratio $\omega_{\rm trig}/\lambda_{\rm eff}$.
%We denote $\gamma$ in this limit as $\gamma_{\rm SI}$.
For $\omega_{\rm trig}/\lambda_{\rm eff}=0$, one recovers the cubic limit $s^x=s^y=s^z=5/6$ and $\gamma=1$.
As $\omega_{\rm trig}/\lambda_{\rm eff}\rightarrow +\infty$, the doublet is dominated by the $|L_{\rm eff}^z=0\rangle$ component, giving $s^{\perp}=1$, $s^z=1/2$, and $\gamma\to 1/4$ (easy-plane).
In contrast, as $\omega_{\rm trig}/\lambda_{\rm eff}\rightarrow -\infty$, the doublet approaches the $|L_{\rm eff}^z=\pm 1\rangle$ sector, yielding $s^{\perp}\to 0$, $s^z\to 3/2$, and $\gamma\to\infty$ (easy-axis).
A numerical evaluation yields the solid curve $\gamma$ as a function of $\omega_{\rm trig}/\lambda_{\rm eff}$ in Fig.~\ref{fig:crystal_alpha}.
For large positive $\omega_{\rm trig}/\lambda_{\rm eff}$, the gap $\Delta E_{\rm ex}$ to the first excited doublet [Fig.~\ref{fig:Co_Crystal} (b)] decreases.
Therefore, the pseudospin-$1/2$ truncation is justified for $J\ll\Delta E_{\rm ex}$.
For {\it XY}CP and {\it X}CSO we find this truncation well satisfied (see Sec.~S5 of SM~\cite{SM}), validating the pseudospin-$1/2$ description.
This establishes a direct link between the local Kramers doublet wavefunction and $\Delta$ as discussed in the followings.

%It should be noted that $\gamma$ is a site parameter, whereas $\Delta$,  which is crucial for the spin supersolid, is a bond parameter. 
We now discuss how the site parameter $\gamma$ controls the bond anisotropy $\Delta$ governing emergence of the spin supersolid.
We consider a total spin exchange tensor $\mathcal{J}_{ij}^{ab}$, acting within the $^{4}T_{1g}$ manifold of Co$^{2+}$ in the absence of $\lambda_{\rm eff}$ and $\omega_{\rm trig}$ [Fig.~\ref{fig:Co_Crystal} (b)].
% (for Co$^{2+}$, the $^{4}T_{1g}$ manifold of high-spin $d^7$).
Upon including $\lambda_{\rm eff}$ and $\omega_{\rm trig}$ and projecting onto the resulting Kramers doublet, one obtains the effective pseudospin-$1/2$ Hamiltonian in Eq.~(\ref{eq:Jani}).
This corresponds to the renormalization $J_{ij}^{ab}=s_i^{a}\mathcal{J}_{ij}^{ab}s_j^{b}$ (see Sec. S4 of SM~\cite{SM}).
Thus, the bond parameter $\Delta$ is controlled by the local single-site parameter $\gamma$ through the projection factors $s_i^{a}$ and $s_j^{b}$.
Since there is only one distinct Co site per unit cell, these projection factors are site independent $s_i^a = s_j^a$, and it satisfies $\Delta \simeq \gamma$~\cite{FN}.
The computed $\Delta$ for the {\it X}BaCP and {\it X}CSO families collapses almost onto the single-ion curve of $\gamma$ as a function of $\omega_{\rm trig}/\lambda_{\rm eff}$ (Fig.~\ref{fig:crystal_alpha}).
This limit then yields a symmetry-protected bound on the exchange anisotropy,
\begin{align}\label{eq:alpha_const}
    \Delta \ge \frac{1}{4},
\end{align}
with $\Delta\to 1/4$ approached for $\omega_{\rm trig}/\lambda_{\rm eff}\to +\infty$ (see Fig.~\ref{fig:crystal_alpha}).
Thus, the ideal XY limit ($\Delta\to 0$) is excluded, whereas the Ising limit ($\Delta\to\infty$) remains allowed.
%The difference between $\Delta$ and $\gamma$ indicates from higher-order crystal field and anisotropic part of $\mathcal{J}_{ij}^{ab}$.
%If this relation is satisfied, $\Delta$ can be represented by a function of $\omega_{\rm trig} / \lambda_{\rm eff}$.
%This is consistent with tiny non XXZ components $J^{\rm PD}$ and $J^{\Gamma}$.
%Therefore, $\Delta$ in these compounds is consistently explained by the difference of trigonal field strength $\omega_{\rm trig}$.
%We note that $\mathcal{J}_{ij}^{{\rm ani},ab}$ is inherently bond-dependent and can be essential for realizing strongly directional interactions, such as Kitaev-type exchange~\cite{Winter_2017}.
%In general, lowering the local symmetry from $D_{3d}$ allows additional crystal-field terms that mix the ground $J_{\rm eff}=1/2$ doublet with higher lying states within the $^{4}T_{1g}$ manifold~\cite{Tanabe-Sugano}. 
%This modifies the projection factors $s^a$ and the range of $\gamma$.
%For the TLAF materials of interest, the local CoO$_6$ environment is well approximated by the trigonal limit discussed above.
%This bound applies to trigonal-field-dressed $J_{\rm eff}=1/2$ Kramers doublet magnets provided the total spin exchange is approximately isotropic $\mathcal{J}_{ij}^{ab}\simeq \mathcal{J}_{ij}^{\rm iso}\delta_{ab}$ and the doublet truncation is justified ($J\ll \Delta E_{\rm ex}$).

The systematic evolution of $\Delta$ by the $X$- and $Y$- site substitution in the {\it XY}CP family and the $X$-site substitution in the {\it X}CSO family are driven predominantly by the O$_6$ contribution (Fig.~\ref{fig:crystal_alpha}).
This indicates that $X$- and $Y$- site substitution acts mainly indirectly, by reshaping the local CoO$_6$ cage, rather than through the direct electrostatic contribution of the $X$- and $Y$-sites themselves.
The structural origin of this oxygen-mediated tuning is the anisotropic lattice relaxation of the quasi-two-dimensional structure.
Because the Co$^{2+}$ ions form a triangular network in the $\mathbf{ab}$ plane, the in-plane framework is more rigid than the interlayer direction, consistent with the negligible interlayer exchange couplings in {\it XY}CP and {\it X}CSO.
As a result, chemical pressure is accommodated mainly by the $\mathbf{c}$-axis response.
For smaller cations, the relaxed structure exhibits a stronger $\mathbf{c}$-axis distortion relative to the $\mathbf{ab}$ plane, which enhances the trigonal elongation of the local CoO$_6$ cage.
In the convention of Eq.~(\ref{eq:ion}), this shifts $\omega_{\rm trig}$ toward more negative values and therefore enhances $\Delta$ through the monotonic relation between $\Delta$ and $\gamma$.
Larger cations produce the opposite structural response, shifting $\omega_{\rm trig}$ upward and reducing $\Delta$.

The point charge analysis reveals the origin of the large difference in the trigonal field strength between {\it X}BaCP and {\it X}CSO families as the cation environment (see Sec.~S6 of SM~\cite{SM}). 
In particular, the large negative trigonal field in {\it X}CSO is dominated by the nearby Se$^{4+}$ ions.
By contrast, in {\it X}BaCP the dominant oxygen and cation contributions partially cancel, yielding a much weaker net trigonal field.

\begin{comment}
We finally examine whether external lattice tuning can push CsCSO further toward smaller $\Delta$.
A simple intuition is provided by the O$^{2-}$ only picture in Fig.~\ref{fig:crystal_alpha}. 
Among local lattice distortions, uniaxial compression along the crystallographic $c$-axis is expected to further reduce $\Delta$.
Motivated by this expectation, we explicitly simulate $c$-axis uniaxial compression in CsCSO.
We find that the compression increases $\omega_{\rm trig}/\lambda$, which in turn reduces $\Delta$ and drives the system toward a less Ising-like regime (Fig.~\ref{fig:CsCSO}).
In particular, our calculations show that about 3\% compression along the $c$-axis already brings $\Delta \sim 5$, near the exact diagonalization boundary for the spin supersolid.
This identifies uniaxial compression as a realistic route to push CsCSO to the spin supersolid.

\begin{figure}[htbp]
\centering
\includegraphics[keepaspectratio, scale=0.4]{./Figs/Fig3.pdf}
\caption{
Effects of uniaxial compression along the crystallographic $c$ axis in CsCSO.
Each point shows the exchange anisotropy $\Delta$ as a function of $\omega_{\rm trig}/\lambda$ for uncompressed CsCSO and under compression.
The labels indicate the $c$-axis strain $(c-c_0)/c_0$ relative to the uncompressed lattice constant $c_0$.
Negative sign indicates compression.}
\label{fig:CsCSO}
\end{figure}
\end{comment}

In summary,  we predict a previously unexplored set of supersolid candidates.
We then show microscopically that the exchange anisotropy $\Delta$ in triangular-lattice Co$^{2+}$ magnets is governed predominantly by the local trigonal crystal field through projection onto the Kramers doublet.
This leads to a material independent relation $\Delta\simeq\gamma$ as a function of $\omega_{\rm trig}/\lambda_{\rm eff}$, with a symmetry protected bound $\Delta\ge 1/4$.
This explains that the $X$- and $Y$-site substitution trends within each family is controlled predominantly by the O$_6$ ligands.
On the other hand, our point charge analysis traces the large difference between the {\it XY}CP and {\it X}CSO families to the cation environment.
These results establish trigonal field engineering as a microscopic design rule for realizing and tuning spin supersolidity in triangular-lattice magnets.

\begin{acknowledgments}
R. O. acknowledges insightful comments from Dr. Igor Solovyev.
R. O. was supported by JSPS KAKENHI Grant No.~JP23KJ2165.
J. I. was supported by JSPS KAKENHI Grant No.~JP24H00409.
M. M. was supported by JSPS KAKENHI Grant No.~JP23K03291 and GIMRT program (No.~202512-QBKNE-0013) of Quantum Beam Center for Materials Research, Institute for Materials Research, Tohoku University.
S. M. was supported by JSPS KAKENHI Grant No.~ 24K00576.
\end{acknowledgments}

\nocite{ONCV,Ballhausen1962,AGeorge,Kanamori_cubic,65bn-63kj,Stevens_1952}
\bibliography{ref}
\end{document}

\title{Supplementary Materials: A microscopic design rule for spin supersolids in triangular-lattice magnets}
\author{Ryota Ono}
\email[Corresponding author: ]{ryota.ono.gm@gmail.com}
\affiliation{Advanced Science Research Center, Japan Atomic Energy Agency, 2-4 Shirakata, Tokai-mura, Ibaraki, 319-1195, Japan}
\author{Jun'ichi Ieda}
\affiliation{Advanced Science Research Center, Japan Atomic Energy Agency, 2-4 Shirakata, Tokai-mura, Ibaraki, 319-1195, Japan}
\author{Michiyasu Mori}
\affiliation{Advanced Science Research Center, Japan Atomic Energy Agency, 2-4 Shirakata, Tokai-mura, Ibaraki, 319-1195, Japan}
\author{Sadamichi Maekawa}
\affiliation{Advanced Science Research Center, Japan Atomic Energy Agency, 2-4 Shirakata, Tokai-mura, Ibaraki, 319-1195, Japan}
\affiliation{RIKEN Center for Emergent Matter Science (CEMS), Wako 351-0198, Japan}

\maketitle

% --- Supplemental numbering ---
\setcounter{secnumdepth}{3} % section番号を出す深さ（念のため）

\makeatletter
\def\@seccntformat#1{\csname the#1\endcsname\quad} % 番号を表示する
\makeatother

\setcounter{section}{0}
\renewcommand{\thesection}{S\arabic{section}}

\setcounter{table}{0}
\renewcommand{\thetable}{S\arabic{table}}

\setcounter{figure}{0}
\renewcommand{\thefigure}{S\arabic{figure}}

\setcounter{equation}{0}
\renewcommand{\theequation}{S\arabic{equation}}

\section{First-principles calculations and low energy Hubbard model}
Electronic-structure calculations were performed using the plane-wave \texttt{Quantum ESPRESSO} package~\cite{QE1} with optimized norm-conserving pseudopotentials~\cite{ONCV}.
Spin-orbit coupling (SOC) is included at the level of fully relativistic pseudopotentials.
We employ the Perdew--Burke--Ernzerhof exchange--correlation functional within the GGA~\cite{GGA}.

In all compounds, the electronic structure near the Fermi level exhibits an isolated manifold of ten bands per Co site
(five $d$ orbitals $\times$ two spin states), consistent with an ionic Co$^{2+}$ picture.
These bands predominantly originate from the Co$^{2+}$ $3d$ orbitals.
This motivates constructing a low-energy Hubbard model by realistic downfolding~\cite{Solovyev_2008_SE,SE_Sergey,PRB_RIS}.
The corresponding 10 low-energy spin-orbitals are constructed using maximally localized Wannier functions implemented in \texttt{wannier90}~\cite{Marzari,Marzari_mod}.
The screened on-site Coulomb interactions are evaluated by the constrained random phase approximation (cRPA) method~\cite{Ferdi}. 

The resulting low-energy effective model reads
\begin{align}
\mathcal{H}^{\rm el}=\mathcal{H}_{\rm one}+\mathcal{H}_U 
\end{align}
where $\mathcal{H}_{\mathrm{one}}$ is the one-electron term and $\mathcal{H}_{\mathrm{U}}$ is the on-site Coulomb interaction.
The one-electron term is
\begin{align}
\mathcal{H}_{\rm one}
=\sum_{ij}\sum_{ab}\sum_{\sigma\sigma'}
h_{ij}^{ab;\sigma\sigma'}\, \hat c_{ia\sigma}^\dagger \hat c_{jb\sigma'},
\end{align}
where $\hat{c}_{ia\sigma}^{\dagger}$ ($\hat{c}_{ia\sigma}$) creates (annihilates) an electron in orbital $a$ with spin $\sigma$ on site $i$.
The on-site block ($i=j$) contains the crystal field and atomic SOC, while the intersite block ($i\neq j$) defines the hopping integrals $t_{ij}^{ab;\sigma\sigma'}\equiv h_{ij}^{ab;\sigma\sigma'}$.

The trigonal field splitting $\omega_{\rm trig}$ is extracted from the on-site multiplet spectrum without including SOC. 
Within the $^4T_{1g}$ manifold, the trigonal crystal field splits the orbital triplet as
$T_{1g}\rightarrow A_{2g}\oplus E_g$. 
In the effective $L_{\rm eff}=1$ representation, $A_{2g}$ corresponds to $L_{\rm eff}^z=0$, while $E_g$ corresponds to $L_{\rm eff}^z=\pm 1$. 
%Because the spin part remains $S=3/2$ in the absence of SOC, the splitting appears as a fourfold $^4A_{2g}$ multiplet and an eightfold $^4E_g$ multiplet. 
Thus, we evaluate
\[
\omega_{\rm trig}=E(^4E_g)-E(^4A_{2g})
\]
directly from the on-site splitting of the $^4T_{1g}$ multiplet.

We also extract the atomic SOC constant $\zeta_d$ by fitting the on-site SOC matrix to the one-electron form $ \frac{\zeta_d}{2}\,\mathbf{l}\cdot\boldsymbol{\sigma} $ where $\mathbf{l}$ is the orbital angular-momentum operator in the $d$-orbital basis and $\boldsymbol{\sigma}$ is the vector of Pauli matrices.
For high-spin $d^7$, the projection of the many-electron SOC operator $\frac{\zeta_d}{2}\sum_i\mathbf l_i\cdot\boldsymbol{\sigma_i}$ onto the parent $^4F$ term gives $
H_{\rm SOC}^{(^4F)}
=
\lambda_{\rm LS}\,\mathbf L\cdot\mathbf S$,
$\lambda_{\rm LS}=-\frac{\zeta_d}{3}$,
where $\mathbf L$ is the true $L=3$ orbital angular momentum.
Within the cubic $^4T_{1g}$ component of the $^4F$ term, the true
orbital operator is related to the fictitious $L_{\rm eff}=1$ operator by $ P_{T_{1g}}\mathbf L P_{T_{1g}}=-\frac32 \,\mathbf L_{\rm eff}$~\cite{Ballhausen1962}.
Therefore the SOC in the $^4T_{1g}$ manifold takes the form$ H_{\rm SOC}^{(^4T_{1g})}=\lambda_{\rm eff}\,\mathbf L_{\rm eff}\cdot\mathbf S$ with $\lambda_{\rm eff}=-\frac32\lambda_{\rm LS}=\frac{\zeta_d}{2}$.
This equality holds in the pure ionic $^4T_{1g}(^4F)$ limit.

The on-site Coulomb interaction $\mathcal{H}_{\rm U}=\sum_i \mathcal{H}_{{\rm U},i}$ is parametrized by the spherical Kanamori form,
\begin{align}
\mathcal{H}_{{\rm U},i}
&= U \sum_a \hat{n}_{ia\uparrow}\hat{n}_{ia\downarrow}
+ U' \sum_{a>b} \hat{n}_{ia}\hat{n}_{ib}
- J_H \sum_{a>b}\sum_{\sigma\sigma'} \hat{c}_{ia\sigma}^\dagger \hat{c}_{ib\sigma'}^\dagger \hat{c}_{ia\sigma'} \hat{c}_{ib\sigma}
+ J_H \sum_{a>b}\left(\hat{c}_{ia\uparrow}^\dagger \hat{c}_{ia\downarrow}^\dagger \hat{c}_{ib\downarrow}\hat{c}_{ib\uparrow} + \text{h.c.}\right),
\end{align}
where $\hat{n}_{ia}=\sum_\sigma \hat{n}_{ia\sigma}$.
All microscopic input parameters of the low-energy Hubbard model and the resulting NN spin-model parameters for $X_2$BaCo(PO$_4$)$_2$ ({\it X}BaCP) and $X_2$Co(SeO$_3$)$_2$ ({\it X}CSO) are summarized in Table~\ref{tab:S1_inputs_outputs}.
The corresponding results for $X_2Y$Co(PO$_4$)$_2$ ({\it XY}CP) are summarized in Table~\ref{tab:XYCP_inputs_outputs}.
The calculated Kanamori parameters are very similar across the compounds and satisfy the spherical-limit relation $U'=U-2J_H$, as expected for $3d$ systems~\cite{AGeorge,Kanamori_cubic}.

\section{Superexchange theory}
In our superexchange calculation we work in the local $d^7$ manifold of Co$^{2+}$.
We construct the many-body basis in the $N=7$ sector built from ten spin-orbitals, so that the local Hilbert-space dimension is $\binom{10}{7}=120$ per site.
To obtain the low-energy degrees of freedom, we diagonalize the on-site Hamiltonian: $\mathcal{H}_{\rm CF}+\mathcal{H}_{\rm SOC}+\mathcal{H}_{\rm U}$ within the $d^7$ sector. 
The lowest-energy multiplet is a Kramers  doublet $\{|\psi_{i+}\rangle,|\psi_{i-}\rangle\}$ at each site $i$.
Next, within this doublet, we calculate maximal/minimal projection onto a pseudo-spin direction ($e_i^x, e_i^y, e_i^z$).
Using these basis states $\left\{\ket{\pm e^a_i}\right\}$ ($a=x,y,z$), we calculate the second-order perturbation energy given by:
\begin{align}
    E_{ij}^{(2)ab} (\mu, \nu)  = \mel{\mu e_i^a, \nu e_j^b}{(\hat{\mathcal{T}}_{ij} + \hat{\mathcal{T}}_{ji})}{\mu e_i^a, \nu e_j^b},
\end{align}
where $\mu,\nu=\pm$, $\ket{\mu e_i^a, \nu e_j^b}=\ket{\mu e_i^a}\otimes \ket{\nu e_j^b}$ is a product state and $\hat{\mathcal{T}}_{ij}$ is a second-order perturbation operator:
\begin{align}
    \hat{\mathcal T}_{ij}= - \sum_{\eta\in(d_i^8,d_j^6)} \frac{\hat{\mathcal P}_{ij}\,\mathcal{H}_{\mathrm{one},ji}\,|d_i^8,d_j^6;\eta\rangle\langle d_i^8,d_j^6;\eta|\,
\mathcal{H}_{\mathrm{one},ij}\hat{\mathcal P}_{ij}}{E_{d_i^8,d_j^6;\eta}-E_0}.
\end{align}
Here $\{|d_i^8,d_j^6;\eta\rangle\}$ denotes a complete set of eigenstates in the virtual charge sector
$(d_i^8,d_j^6)$ with energies $E_{d_i^8,d_j^6;\eta}$.
$\hat{\mathcal P}_{ij}$ is a projector excluding unphysical states. 
$E_0$ is the energy of the initial $(d_i^7,d_j^7)$ state.
Another virtual $(d_i^6,d_j^8)$ sector is included in $ \hat{\mathcal T}_{ji}$.
After obtaining the $E_{ij}^{(2)ab} (\mu, \nu)$, we can map onto an anisotropic exchange tensor by $J_{ij}^{ab} =( E_{ij}^{(2)ab} (+, +) + E_{ij}^{(2)ab} (-, -) - E_{ij}^{(2)ab} (+, -) - E_{ij}^{(2)ab} (-, +))/4$.
Using these procedures, we obtain exchange parameters for all target compounds listed in Table~\ref{tab:S1_inputs_outputs}.

\begin{comment}
\section{Validity of $\Delta \approx \gamma$}
As it was stated in the main text, equality between $\Delta$ and $\gamma$ depends on the anisotropy in the spin exchange $\tilde{J}_{ij}^{ab}$.
Predicting this inequality can be done by checking the SU(2)-likeness of the hopping integral.
We first introduce the decomposition of the hopping matrix $T_{ij}$ by the Pauli matrices as 
\begin{align}
    T_{ij} = T_{ij}^0 \mathbb{I}_2 + \sum_\Delta  T_{ij}^a \sigma^{a},
\end{align}
where $T_{ij}^0$, $T_{ij}^a$ are decomposed hopping matrix into SU(2) part and the SOC dependent part, respectively. 
\end{comment}

\begin{table*}[htpb]
\caption{Microscopic input parameters of the low-energy Hubbard model and the resulting NN spin-model parameters for
$X_2$BaCo(PO$_4$)$_2$ ({\it X}BaCP) and $X_2$Co(SeO$_3$)$_2$ ({\it X}CSO), $X=$ Na, K, Rb, Cs.
Values in parentheses are from experimental estimates~\cite{PNAS_NBCP,Gao2022,KCSO_npj,65bn-63kj}.
We use the convention $\omega_{\rm trig}=E(^4E_g)-E(^4A_{2g})$ and $\lambda_{\rm eff}=\zeta_d/2$ for high-spin $d^7$.}
\label{tab:S1_inputs_outputs}
\begin{ruledtabular}
\setlength{\tabcolsep}{2.6pt} % (optional) tighten a bit since we add 2 columns
\renewcommand{\arraystretch}{0.95}
\begin{tabular}{c ccc c cc ccccc}
$X$
& $U$ & $U'$ & $J_H$
& $\zeta_d$
& $\omega_{\rm trig}$
& $\omega_{\rm trig}/\lambda_{\rm eff}$
& $J$ & $\Delta$ & $J^{\rm PD}$ & $J^\Gamma$ & $\gamma$ \\
& (eV) & (eV) & (eV) & (meV) & (meV) &  & (meV) &  & (meV) & (meV) &  \\
\hline
\multicolumn{12}{l}{$X_2$BaCo(PO$_4$)$_2$ ({\it X}BaCP)}\\
Na & 5.83 & 4.35 & 0.74 & 68 (66) & -16 (-11) & -0.47  & 0.132\,(0.076) & 1.1 (1.7) & -0.010 & 0.006 & 1.20 \\
K  & 5.26 & 3.80 & 0.73 & 67 & -1   & -0.03  & 0.127          & 0.8        & -0.009 & -0.006 & 0.85 \\
Rb & 5.02 & 3.58 & 0.73 & 66 & +7   & +0.22  & 0.123          & 0.6        & -0.009 & -0.005 & 0.71 \\
Cs & 4.69 & 3.28 & 0.71 & 68 & +22  & +0.64  & 0.118          & 0.4        & -0.010 & -0.003 & 0.49 \\
\hline
\multicolumn{12}{l}{$X_2$Co(SeO$_3$)$_2$ ({\it X}CSO)}\\
Na & 5.18 & 3.72 & 0.74 & 70 & -255 & -7.29 & 0.056          & 33.3       & 0.000  & 0.003  & 34.90 \\
K  & 5.40 & 3.93 & 0.74 & 70 & -208 & -5.94  & 0.084\,(0.217) & 22.6 (14.3) & 0.000  & 0.005  & 23.62 \\
Rb & 5.27 & 3.79 & 0.75 & 67 & -176 & -5.25  & 0.126          & 13.2       & -0.001 & 0.007  & 13.70 \\
Cs & 5.18 & 3.70 & 0.74 & 66 & -146 & -4.42  & 0.203          & 7.4        & -0.001 & 0.010  & 7.63 \\
\end{tabular}
\end{ruledtabular}
\end{table*}

\begin{table*}[htpb]
\caption{Microscopic input parameters of the low-energy Hubbard model and resulting NN spin-model parameters for $X_2Y$Co(PO$_4$)$_2$ ({\it XY}CP), with $X=$ Na, K, Rb, Cs and $Y=$ Mg, Ca, Sr, Ba. We use $\omega_{\rm trig}=E(^4E_g)-E(^4A_{2g})$ and $\lambda_{\rm eff}=\zeta_d/2$ for high-spin $d^7$.
The Kanamori interaction parameters $U$, $U'$, and $J_H$ are fixed to those of the corresponding parent
$X$BaCP compound.}
\label{tab:XYCP_inputs_outputs}
\begin{ruledtabular}
\scriptsize
\setlength{\tabcolsep}{2.2pt}
\renewcommand{\arraystretch}{0.95}
\begin{tabular}{cc ccc ccc ccccc}
$X$ & $Y$ & $U$ & $U'$ & $J_H$ & $\zeta_d$ & $\omega_{\rm trig}$ & $\omega_{\rm trig}/\lambda_{\rm eff}$ & $J$ & $\Delta$ & $J^{\rm PD}$ & $J^\Gamma$ & $\gamma$ \\
& & (eV) & (eV) & (eV) & (meV) & (meV) & & (meV) & & (meV) & (meV) & \\
\hline
\multirow{4}{*}{Na}
& Mg & \multirow{4}{*}{5.83} & \multirow{4}{*}{4.35} & \multirow{4}{*}{0.74}
& 67.2 & -121.8 & -3.62 & 0.033 & 8.60 & 0.001 & -0.005 & 10.00 \\
& Ca &  &  &  & 67.8 & -61.2 & -1.81 & 0.077 & 3.00 & 0.003 & 0.007 & 3.34 \\
& Sr &  &  &  & 67.1 & -41.1 & -1.23 & 0.104 & 1.87 & 0.006 & 0.006 & 2.06 \\
& Ba &  &  &  & 68.2 & -16.0 & -0.47 & 0.132 & 1.07 & 0.010 & 0.006 & 1.19 \\
\hline
\multirow{4}{*}{K}
& Mg & \multirow{4}{*}{5.26} & \multirow{4}{*}{3.80} & \multirow{4}{*}{0.73}
& 66.4 & -95.6 & -2.88 & 0.034 & 6.49 & 0.001 & -0.005 & 7.22 \\
& Ca &  &  &  & 67.0 & -37.0 & -1.10 & 0.080 & 1.93 & 0.004 & -0.006 & 2.10 \\
& Sr &  &  &  & 66.4 & -18.2 & -0.55 & 0.105 & 1.17 & 0.007 & -0.005 & 1.28 \\
& Ba &  &  &  & 67.2 & -0.8 & -0.02 & 0.127 & 0.77 & 0.009 & -0.006 & 0.85 \\
\hline
\multirow{4}{*}{Rb}
& Mg & \multirow{4}{*}{5.02} & \multirow{4}{*}{3.58} & \multirow{4}{*}{0.73}
& 65.7 & -92.0 & -2.80 & 0.032 & 5.73 & 0.001 & -0.005 & 6.15 \\
& Ca &  &  &  & 66.3 & -24.4 & -0.74 & 0.077 & 1.43 & 0.004 & 0.006 & 1.55 \\
& Sr &  &  &  & 65.7 & -5.6 & -0.17 & 0.099 & 0.86 & 0.007 & -0.004 & 0.95 \\
& Ba &  &  &  & 66.8 & 6.6 & 0.20 & 0.122 & 0.63 & 0.009 & -0.005 & 0.71 \\
\hline
\multirow{4}{*}{Cs}
& Mg & \multirow{4}{*}{4.69} & \multirow{4}{*}{3.28} & \multirow{4}{*}{0.71}
& 65.0 & -67.2 & -2.07 & 0.027 & 4.88 & 0.001 & -0.005 & 5.16 \\
& Ca &  &  &  & 64.7 & -12.1 & -0.37 & 0.066 & 1.12 & 0.003 & -0.006 & 1.21 \\
& Sr &  &  &  & 65.3 & 14.5 & 0.44 & 0.099 & 0.44 & 0.007 & -0.004 & 0.53 \\
& Ba &  &  &  & 65.7 & 21.7 & 0.66 & 0.117 & 0.41 & 0.010 & -0.003 & 0.49 \\
\end{tabular}
\end{ruledtabular}

\end{table*}

\section{Exchange interactions further than nearest-neighbor}
In general, exchange interactions are not restricted to the nearest-neighbor (NN) bond.
We therefore also examine further-neighbor couplings.
Here $J_2$ and $J_3$ denote the second- and third-nearest-neighbor exchanges, respectively.
In our structural geometry, the second-neighbor bond $J_2$ connects adjacent triangular layers and thus provides the leading interlayer coupling.
This coupling is crucial for assessing whether a purely two-dimensional triangular-lattice antiferromagnet (TLAF) model is a good approximation.
Table~\ref{tab:Jfar} summarizes the exchanges from $J_2$ (second NN) to $J_3$ (third NN).
For simplicity, we show only the isotropic component,
$J_n^{\rm iso}\equiv \frac{1}{3}\sum_{a=x,y,z}J_n^{aa}$.

As shown in Table~\ref{tab:Jfar}, the further-neighbor exchanges are substantially smaller than the NN interaction.
For the {\it X}BaCP family, $J_2/J_1\simeq(0.1\text{--}0.25)\%$ and $J_3/J_1\simeq(0.02\text{--}0.14)\%$.
For the {\it X}CSO family, $J_3$ is negligible ($J_3/J_1\lesssim 5\times10^{-4}$), while $J_2$ remains subdominant with $J_2/J_1\le 5.1\%$ (maximum at NaCSO).
These results support a minimal TLAF description dominated by $J_1$, with only weak corrections from further-neighbor couplings.

We also observe a systematic reduction of $J_2$ and $J_3$ upon substituting the {\it X}-site ion from Na to Cs,
This is consistent with lattice expansion for larger {\it X} ions, which suppresses orbital overlap and hence superexchange.

\begin{table*}[htpb]
\caption{Isotropic exchange interactions $J_n^{\rm iso}$ (meV) for the NN ($J_1^{\rm iso}$), second-NN ($J_2^{\rm iso}$), and third-NN ($J_3^{\rm iso}$) bonds.}
\label{tab:Jfar}
\begin{ruledtabular}
\begin{tabular}{c ccc}
$X$ & $J_1^{\rm iso}$ & $J_2^{\rm iso}$ & $J_3^{\rm iso}$  \\
\hline
\multicolumn{4}{l}{$X_2$BaCo(PO$_4$)$_2$ ({\it X}BaCP)}\\
Na & 0.13516 & 0.00030 & 0.00019   \\
K  & 0.11721 & 0.00029 & 0.00007   \\
Rb & 0.10761 & 0.00025 & 0.00006   \\
Cs & 0.09440 & 0.00010 & 0.00002   \\
\hline
\multicolumn{4}{l}{$X_2$Co(SeO$_3$)$_2$ ({\it X}CSO)}\\
Na & 0.65972 & 0.03369 & 0.00010   \\
K  & 0.68933 & 0.01097 & 0.00033   \\
Rb & 0.63592 & 0.00493 & 0.00008   \\
Cs & 0.63635 & 0.00154 & 0.00002   \\
\end{tabular}
\end{ruledtabular}
\end{table*}

\section{Single-ion spin-orbit entanglement in Kramers doublet}
%To understand the microscopic origin of the exchange anisotropy, we first isolate the single-ion contribution in a generic Kramers doublet system.
Let $\{|\psi_{i+}\rangle,|\psi_{i-}\rangle\}$ be the local ground state Kramers doublet on site $i$, obtained by diagonalizing the on-site Hamiltonian in many-body basis. 
Projecting the total spin operator $S_i^a$ onto this doublet yields a $2\times2$ matrix,
\begin{align}
&\mel{\psi_{is}}{S_i^a}{\psi_{is'}}
= \frac{1}{2}\sum_{\mu=x,y,z}\Lambda_i^{a\mu}\,(\sigma^\mu)_{ss'},\\
&s,s'=\pm,\ a=x,y,z, \nonumber
\end{align}
where $\sigma^\mu$ are the Pauli matrices in the $\{|\psi_{i+}\rangle,|\psi_{i-}\rangle\}$ basis.
The real $3\times3$ matrix $\Lambda_i^{a \mu}$ encodes the single-ion spin-orbit entanglement.
Then, the maximal projection along each pseudospin direction is given by $ s^{a}_i \equiv \sqrt{\sum_{\mu=x,y,z}(\Lambda_i^{a \mu})^2}$.
In the presence of local $D_{3d}$ symmetry as in the case of {\it X}BaCP and {\it X}CSO, one can choose the pseudospin axes such that $S_i^{x,y}=s_i^{\perp}\, e_i^{x,y}$ and $S_i^{z}=s_i^{z}\, e_i^{z}$, with $s_i^{x}=s_i^{y}\equiv s_i^\perp$.

\section{Validity of the Kramers-doublet (pseudospin-$1/2$) description}

We evaluate the realistic splitting $\Delta E_{\rm ex}=E_1-E_0$ by exact diagonalization of the full on-site Hamiltonian
$\mathcal{H}_{\rm CF}+\mathcal{H}_{\rm SOC}+\mathcal{H}_{\rm U}$ in the local $d^7$ sector.
Table~\ref{tab:Kramers} shows that $\Delta E_{\rm ex}\simeq 20$--$32$~meV for all target materials,
which is much larger than the exchange scale (of order meV or smaller).
This confirms that the low-energy physics is well captured by the Kramers-doublet (pseudospin-$1/2$) description.

\begin{table}[htpb]
\centering
\caption{
Energy splitting between the ground state Kramers doublet ($E_0$) and a next lowest multiplet ($E_1$): $\Delta E_{\rm ex} = E_1-E_0$ in {\it X}BaCP and {\it X}CSO (meV).
A value in parenthesis is an experimental splitting from X-ray absorption spectroscopy~\cite{65bn-63kj}.
}
\label{tab:Kramers}
\begin{ruledtabular}
\begin{tabular}{c cccc cccc}

& NaBaCP & KBaCP & RbBaCP & CsBaCP & NaCSO & KCSO & RbCSO & CsCSO\\
\hline
$\Delta E_{\rm ex}$ & 29.93 (40) & 31.56 & 29.19 & 23.52 & 20.07 & 20.45 & 19.50 & 19.88
\end{tabular}
\end{ruledtabular}
\end{table}

\begin{figure}[htbp]
\centering
\includegraphics[keepaspectratio, scale=0.4]{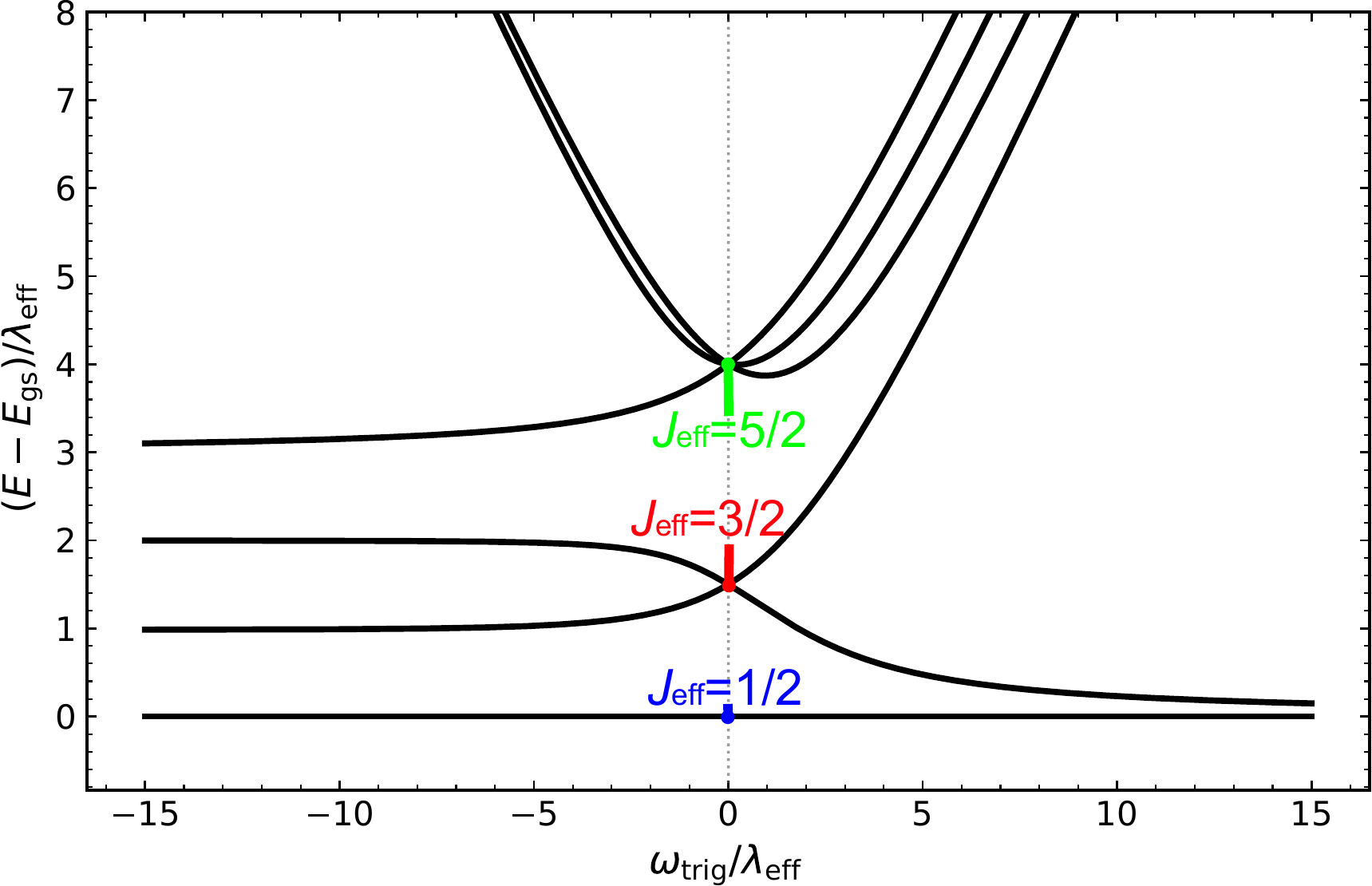}
\caption{
Energies of the two lowest Kramers doublets, measured relative to the ground-state energy, as a function of $\omega_{\rm trig}/\lambda$. 
For $\omega_{\rm trig}\neq 0$, $J$ is not strictly conserved. 
The levels are therefore labelled by the dominant $J_z$ character.
}\label{fig:doublets}
\end{figure}

\section{point charge crystal field analysis}
Here, we analyze the crystal field effect at Co$^{2+}$ in {\it X}BaCP and {\it X}CSO families.
We start from the point charge approximation for the electric potential at $\mathbf{r}$:
\begin{align}
    V(\mathbf{r}) = \sum_{\alpha} \frac{q_\alpha}{|\mathbf{R}_\alpha-\mathbf{r}|},
\end{align}
where $q_\alpha$ is an effective ionic charge of ion $\alpha$ located at $\mathbf R_\alpha$ (with respect to the Co site taken as the origin).

For $|\mathbf r|\ll |\mathbf R_\alpha|$, the Coulomb kernel can be expanded in spherical harmonics as
\begin{align}
\frac{1}{|\mathbf R-\mathbf r|}
=\sum_{k=0}^\infty \sum_{q=-k}^{k}
\frac{4\pi}{2k+1}\frac{r^k}{R^{k+1}}
Y_{kq}(\Omega_{\mathbf r})\,Y_{kq}^*(\Omega_{\mathbf R}),
\end{align}
where $\Omega_{\mathbf r}$ and $\Omega_{\mathbf R}$ denote the solid angles of $\mathbf r$ and $\mathbf R$, respectively.
%(See, e.g., Ref.~\cite{HUTCHINGS1964227} for its use in point charge crystal field calculations).

Choosing the local trigonal axis $\hat{\mathbf n}$ (defined from the local CoO$_6$ geometry), we focus on the axial ($q=0$) components of the lattice potential.
Their angular dependence can be decomposed with respect to Legendre polynomials as $Y_{k0}(\Omega)\propto P_k(\cos\theta)$ with $\cos\theta=\hat{\mathbf n}\cdot \hat{\mathbf R}$.
This motivates defining the axial multipole lattice sums
\begin{align}
\mathcal{A}_{k0}\ \equiv\ \sum_\alpha q_\alpha\,\frac{P_k(\cos\theta_\alpha)}{R_\alpha^{k+1}},
\qquad
\cos\theta_\alpha=\hat{\mathbf n}\cdot \hat{\mathbf R}_\alpha ,
\end{align}
up to an overall prefactor common to all materials.
Within the Stevens operator equivalent formalism~\cite{Stevens_1952},
the axial crystal field parameters satisfy $B_k^{0}\propto A_{k0}\langle r^k\rangle$, so $\mathcal{A}_{k0}$ serves as a convenient geometric decomposition of the axial crystal field.
For $d$ electrons ($l=2$), the leading nontrivial axial terms are $k=2$ and $k=4$ (the $k=0$ term is a constant shift).
Using $P_2(x)=(3x^2-1)/2$ and $P_4(x)=(35x^4-30x^2+3)/8$, we obtain
\begin{align}
\mathcal{A}_{20}
&=\sum_\alpha q_\alpha\,\frac{P_2(\cos\theta_\alpha)}{R_\alpha^{3}}
=\frac12\sum_\alpha q_\alpha\,\frac{3\cos^2\theta_\alpha-1}{R_\alpha^{3}},
\\
\mathcal{A}_{40}
&=\sum_\alpha q_\alpha\,\frac{P_4(\cos\theta_\alpha)}{R_\alpha^{5}}
=\frac18\sum_\alpha q_\alpha\,\frac{35\cos^4\theta_\alpha-30\cos^2\theta_\alpha+3}{R_\alpha^{5}}.
\end{align}
In practice, we use $\mathcal{A}_{20}$ and $\mathcal{A}_{40}$ as lattice sum descriptors to identify which coordination shells and chemical units dominantly generate the trigonal crystal field.

\begin{figure}[htbp]
\centering
\includegraphics[keepaspectratio, scale=0.18]{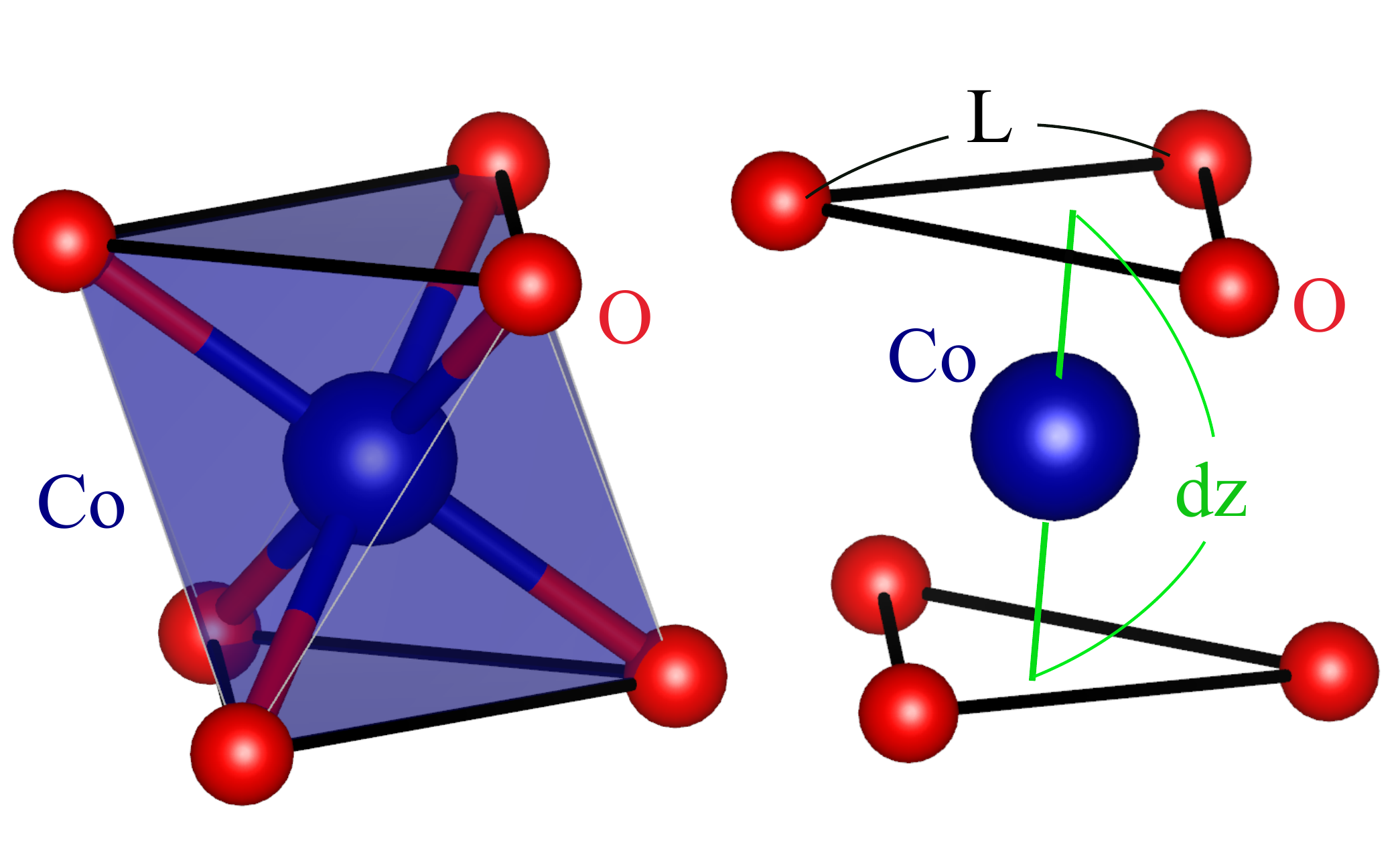}
\caption{ 
Local CoO$_6$ geometry. Crystallographic measures used to determine $r=\sqrt{3/2} (dz/ L)$ are indicated as black solid lines and a green solid line, respectively.
} \label{fig:O6}
\end{figure}

We define $L$ as the side length of each O$_3$ triangle and $d_z$ as the separation between the two O$_3$ triangle planes along the local trigonal axis, as shown in Fig.~\ref{fig:O6}.
We then introduce the dimensionless trigonal distortion parameter $r=\sqrt{3/2} (dz/ L)$.
For an ideal octahedron, this definition gives $r=1$.
Thus, $r>1$ corresponds to a trigonal elongation of the O$_6$ cage along the local trigonal axis, whereas $r<1$ corresponds to a trigonal compression.
This provides a simple local structural descriptor for the ligand contribution to the trigonal crystal field.
However, $r$ characterizes only the nearest-neighbor O$_6$ geometry and does not by itself determine the full crystal field, because the axial lattice sums $\mathcal{A}_{20}$ and $\mathcal{A}_{40}$ also contain contributions from more distant cations through their charges and distances.
Therefore, in the following we use $r$ to diagnose the local O$_6$ distortion, while $\mathcal{A}_{20}$ and $\mathcal{A}_{40}$ are used to decompose the total point charge crystal field into ligand and cation contributions.

\begin{figure}[htbp]
\centering
\includegraphics[keepaspectratio, scale=0.35]{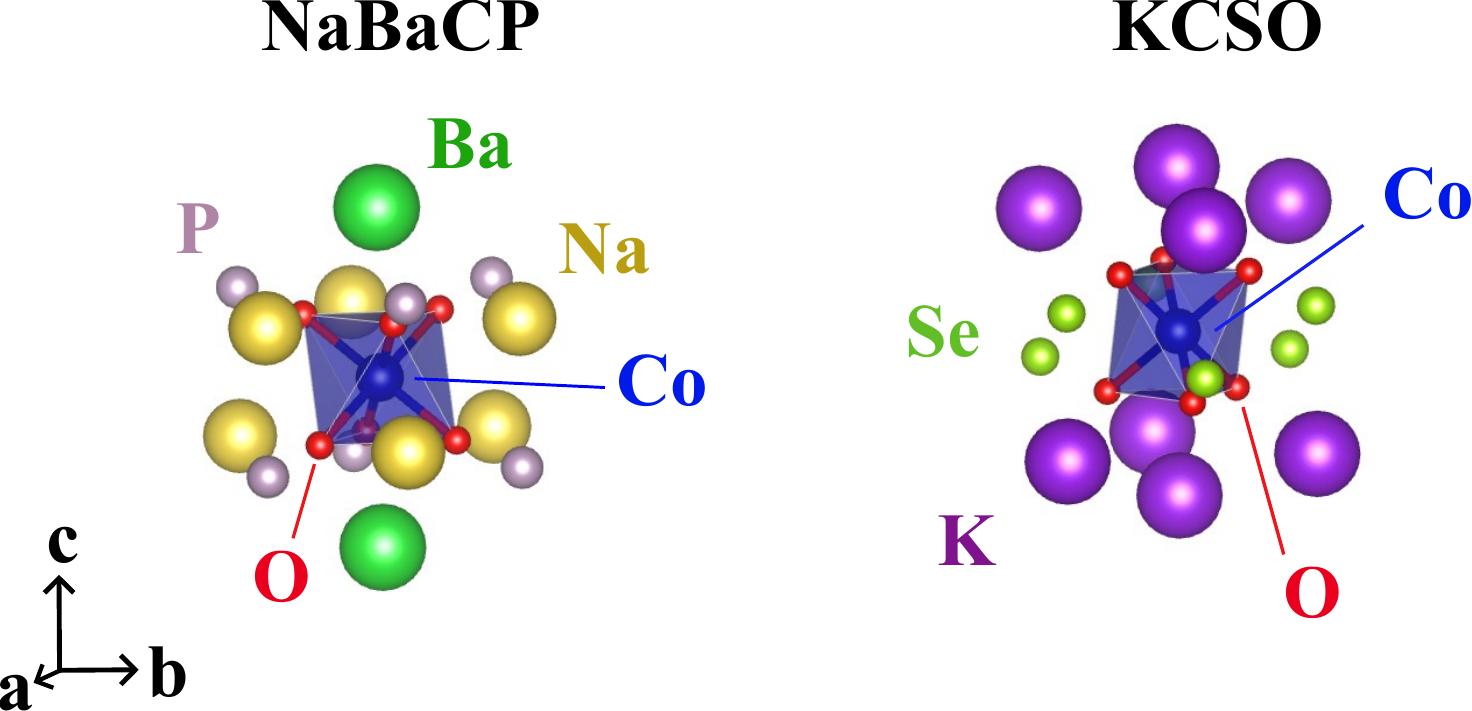}
\caption{ 
Local geometry around a CoO$_6$ cluster in NaBaCP and KCSO. 
} \label{fig:O6_loc}
\end{figure}

\begin{figure}[htbp]
\centering
\includegraphics[keepaspectratio, scale=0.6]{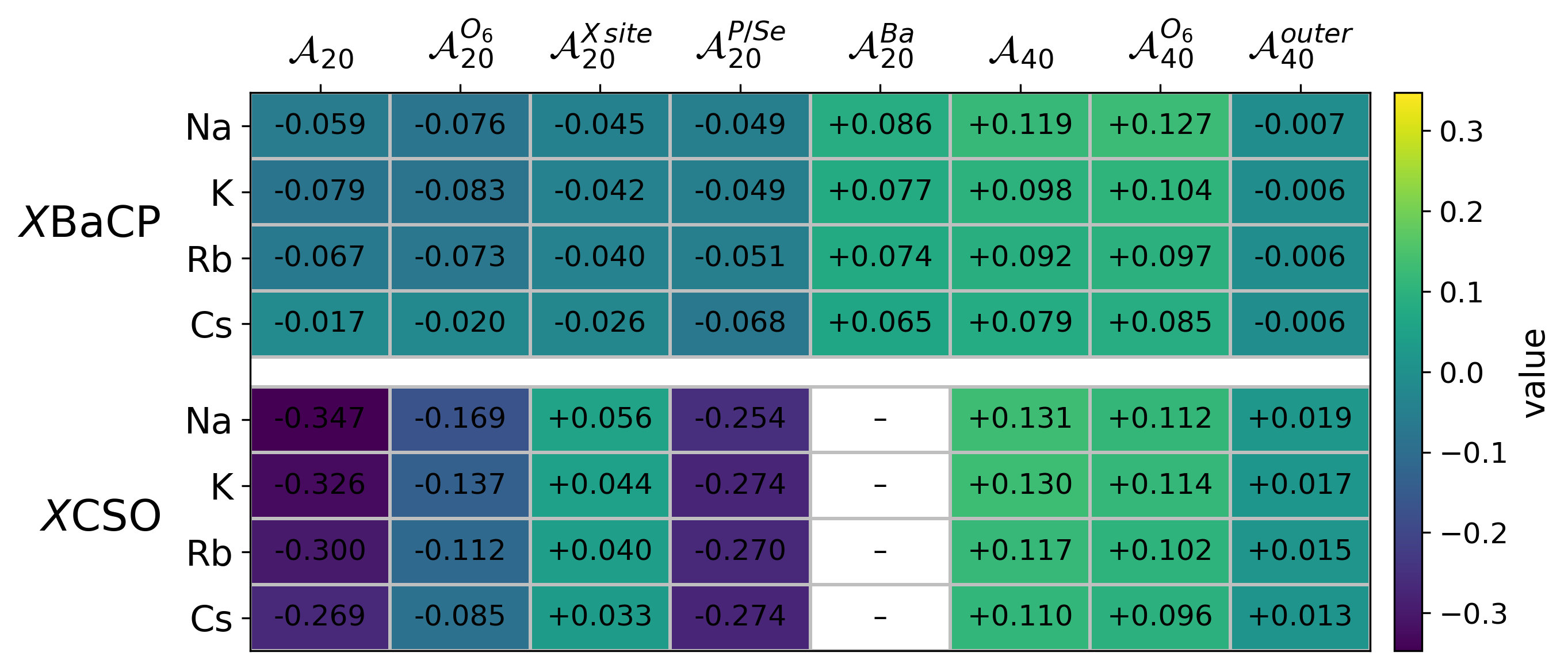}
\caption{
Results of the point charge lattice sum analysis, using the definitions with Legendre prefactors $P_2$ and $P_4$ for {\it X}BaCP and {\it X}CSO.
We list the axial multipole sums $\mathcal{A}_{20},\mathcal{A}_{40}$ decomposed into the nearest O$_6$ and the outer environment, together with species resolved contributions to $\mathcal{A}_{20}$.
For {\it X}BaCP we additionally list the Ba contribution to $\mathcal{A}_{20}$.
Overall common prefactors are omitted.
Color indicates the sign and magnitude of each contribution.
} \label{tab:CEF}
\end{figure}

The results of the axial multipole lattice sums for {\it X}BaCP and {\it X}CSO are summarized in Fig.~\ref{tab:CEF}.
The dominant contrast between the two families appears in the quadrupolar component $\mathcal{A}_{20}$.
$|\mathcal{A}_{20}|$ is strongly suppressed in {\it X}BaCP, whereas it remains large and negative in {\it X}CSO.
This large difference mainly originates from the difference in the outer cation environment.
In {\it X}CSO, the outer environment contribution $\mathcal{A}_{20}^{\rm outer}$ is sizeable and is dominated by the Se$^{4+}$ units, yielding a large negative contribution that adds to the local O$_6$ component.
In contrast, in {\it X}BaCP, the local O$_6$ contribution $\mathcal{A}_{20}^{O_6}$ is partially compensated by positive contributions from the surrounding cations, most notably Ba$^{2+}$, resulting in a near cancellation of the total $\mathcal{A}_{20}$.
We also show the local environment around CoO$_6$ in NaBaCP and KCSO in Fig.~\ref{fig:O6_loc}.
This tendency is captured by this local environment.
In {\it X}CSO, Se$^{4+}$ is located almost in the triangular-lattice plane.
In contrast, in {\it X}BaCP, P$^{5+}$ and the $X^{+}$ sites are located above and below the triangular-lattice plane.
This difference is crucial for the contrast between the cation contributions: $\mathcal{A}_{20}^{P}$ and $\mathcal{A}_{20}^{X{\rm site}}$ in {\it X}BaCP, and $\mathcal{A}_{20}^{\rm Se}$ in {\it X}CSO.

%Notably, $\mathcal{A}_{20}^{\rm outer}$ is large and negative in {\it X}CSO but small and positive in {\it X}BaCP, indicating that the outer chemical units enhance the trigonal field only in the Se-based family.

Furthermore, upon increasing the alkali-metal ionic radius from Na to Cs, the magnitude of the local O$_6$ contributions systematically decreases.
This systematic change is also reflected in the O$_6$ trigonal distortion parameter $r$ (Table~\ref{tab:s}).
This trend can be attributed to structural relaxation of the O$_6$ ligand cage.
For smaller alkali-metal ions, the surrounding cage has more room to relax, which enhances the trigonal distortion of O$_6$ and results in a larger $|\mathcal{A}_{20}^{O_6}|$.

\begin{comment}
The results of the axial multipole lattice sums for {\it X}BaCP and {\it X}CSO are summarized in Table~\ref{tab:CEF}.
It is clear that the huge difference between ABaCP and ACSO families is coming from difference in quadrupolar component $\mathcal{A}_{20}$. 
In ACSO, the $\mathcal{A}_{20}$ from Se$^{4+}$ and O$^{2-}$ are the same sign, while in ABaCP,  $\mathcal{A}_{20}$ from Ba$^{+}$ and O$^{2-}$ are opposite sign. 
As a result, total $\mathcal{A}_{20}$ almost vanishes in ABaCP, while largely survive in ACSO.
In addition, the tendency of Alkaline-metal substitution (from Na to Cs) is overall reduction of the $\mathcal{A}_{20}$ and the $\mathcal{A}_{40}$ from O$_6$.
This is due to the lattice expansion caused by the larger ionic radius of the larger $A$ ion.
\end{comment}

\begin{table}[htpb]
\centering
\caption{Trigonal distortion parameter $r$ for {\it X}BaCP and {\it X}CSO.
Experimental structural values are shown in parentheses, while the other values are obtained from theoretical structural relaxation.}
\begin{tabular}{lcc}
\hline\hline
$X$ & {\it X}BaCP & {\it X}CSO \\
\hline
Na & (1.088) & 1.220 \\
K  & 1.109   & (1.175) \\
Rb & 1.099   & 1.150 \\
Cs & 1.030   & 1.116 \\
\hline\hline
\end{tabular}
\label{tab:s}
\end{table}

\bibliography{ref_sup}